\documentclass[draft]{agujournal}
\draftfalse

\journalname{JGR-Atmospheres}

\begin{document}

%
%

\title{Characterizing upward lightning with and without a terrestrial gamma-ray flash}

%
%




\authors{D. M. Smith\affil{1}, G. S. Bowers\affil{2},
M. Kamogawa\affil{3}, D. Wang\affil{4}, T. Ushio\affil{5}  J. Ortberg\affil{1},
J. R. Dwyer\affil{6}, and M. Stock\affil{7}}

\affiliation{1}{Physics Department and Santa Cruz Institute for Particle Physics, University of California, Santa Cruz, California, USA}
\affiliation{2}{Los Alamos National Laboratory, Los Alamos, California, USA}
\affiliation{3}{Tokyo Gakugei University, Tokyo, Japan}
\affiliation{4}{Gifu University, Gifu, Japan}
\affiliation{5}{Tokyo Metropolitan University, Tokyo, Japan}
\affiliation{6}{University of New Hampshire, Durham, New Hampshire, USA}
\affiliation{7}{Earth Networks, Germantown, MD, USA }




\correspondingauthor{David M. Smith}{dsmith8@ucsc.edu}




\begin{keypoints}
\item Similar upward leaders approaching a thundercloud may
or may not produce a terrestrial gamma-ray flash (TGF).
\item The upper limit on gamma-ray fluence from one flash with
no TGF is $10^7$ times lower than the inferred TGF fluence from a similar flash.
\item Only the TGF-producing leader happened during a gamma-ray glow, implying a higher pre-existing in-cloud field when TGFs occur.
\end{keypoints}

%
%


\begin{abstract}
We compare two observations of gamma-rays before, during, and after lightning flashes
initiated by upward leaders from a tower during low-altitude winter thunderstorms on the
western coast of Honshu, Japan.  While the two leaders appear similar, one produced a terrestrial
gamma-ray flash (TGF) so bright that it paralyzed the gamma-ray detectors while it was occurring,
and could be observed only via the weaker flux of neutrons created in its wake, while the
other produced no detectable TGF gamma-rays at all.  The ratio between the indirectly derived
gamma-ray fluence for the TGF and the 95\% confidence gamma-ray upper limit for the gamma-ray quiet flash is a factor of $1.0 \times 10^7$.  With the only two observations of this type providing such dramatically
different results -- a TGF probably as bright as those seen from space and a powerful upper limit --
we recognize that weak, sub-luminous TGFs in this situation are probably not common, and we quantify this conclusion.  While the gamma-ray quiet flash appeared to have a faster leader and more powerful initial
continuous current pulse than the flash that produced a TGF, the TGF-producing flash
occurred during a weak gamma-ray "glow", while the gamma-ray quiet flash did not, implying a 
higher electric field aloft when the TGF was produced.  We suggest that the field in the
high-field region approached by a leader may be more important for whether a TGF is produced than
the characteristics of the leader itself.  
\end{abstract}

%
%

%


%
%
%
%

\section{Introduction}

Bright, brief flashes of gamma radiation have been observed in association with lightning that initiates with upward leaders, both from tall structures in winter thunderstorms in Japan \citep{bowers17} and from the top of the wire in wire-triggered flashes in Florida \citep{dwyerground, hare16}.  These events may be as intrinsically bright \citep{bowers17} as the terrestrial gamma-ray flashes (TGFs) seen from space \citep{fishSci}. 
That these events have a similar gamma-ray energy spectrum to TGFs -- extending beyond 5--10~MeV -- has been demonstrated directly, if with limited photon statistics, by \citet{dwyerground}, and indirectly by the copious production of photoneutrons -- which require gamma-rays above $\sim10$~MeV \citep{babich07neutrons,babich08neutrons,carlson10neutron} -- in those events produced close enough to the ground for the neutrons to propagate to ground-based detectors \citep[][the latter event having been produced by a flash not yet determined to have come from an upward or downward leader]{bowers17,enoto17}.  

In this paper, we will interpret these downward TGFs as being related to the interaction of lightning leaders with a region of 
pre-existing high electric field
already producing a "glow" of enhanced gamma-ray emission, possibly due to 
relativistic runaway electron avalanches (RREA); a discussion of all these phenomena is therefore a necessary introduction.

Multiple brief ($\sim1$ microsecond) bursts of lower-energy x-rays (up to one or at most a few MeV) are often seen associated with descending negative stepped or dart leaders \citep[e.g.][]{moore01,dwyerLeader05,saleh09,schaal12,montanya14} in both natural and triggered lightning.  These are interpreted as due to "cold runaway", in which a very high field exists in a small region at the leader tip during stepping, accelerating every available free electron to semi-relativistic energies.  X-ray observations give estimates of the 
number of such electrons produced in a step in the range of $10^{11}$ to
$10^{12}$ \citep{dwyer10dose,schaal12}.  

This cold runaway process and its resulting spectrum stands in contrast to the RREA process thought to give TGFs their characteristic brightness and specific, harder energy spectrum. In RREA, a lower electric field extends over a much larger distance, producing a potential on the order of 50~MeV or more.  Electrons that are already semi-relativistic (typically with keV energies) experience low enough friction to continue accelerating to energies approaching the total potential, while lower-energy electrons experience higher friction and slow to a low terminal velocity \citep{wilson}.  On rare occasions, the fast electrons undergo M\o ller scattering and eject an electron from an atom at high enough energy that it too can run away; this produces a relativistic avalanche in which only a small fraction of all free electrons participate \citep{gurev92}.  

Because TGFs are too bright to result from RREA acting on the ambient population of energetic electrons from cosmic rays, there are three leading models to produce TGFs in association with lightning: in the first two, the energetic electrons already known to be associated with stepped leaders, probably originating in the streamer zone ahead of them, serve as seed electrons that undergo RREA in a weaker field.  This subsequent field region can be provided either by the leader itself \citep{moss06,carlson10,celestin11}, or it can be the pre-existing, ambient thunderstorm field
\citep{moss06,dwyer08,dwyer12tgftheory}.  In the third class of models, seeds from the leader are not needed; instead, each avalanche breeds further avalanches in a feedback
process where new seeds are created by positrons and gamma-rays returning to the start of the avalanche region \citep{dwyer03limit,babich05,dwyer08,dwyer12tgftheory}.

The clean conceptual separation of (weak, soft) lightning stepped leader x-rays and (bright, hard) TGFs is blurred by the existence of observations that might be considered intermediate, such as relatively hard but faint events during downward leaders \citep{abbasi17,belz18}, just after a cloud-to-ground return stroke \citep{dwyer12return}, and with lightning generally \citep{ringuette13}.  A continuum of behaviors between soft stepped-leader emissions and TGFs can be considered, with either a higher-potential sort of cold runaway in particularly powerful leaders or situations where there is a limited amount of relativistic runaway in addition to cold runaway.  At the same time, the temporal difference between "classical" stepped leader emission, in discrete microsecond pulses, and TGFs, with gamma-rays emitted continuously for tens to hundreds of microseconds, should be kept in mind. As the bright downward events 
observed by \citet{dwyerground} and \citet{hare16} show, the continuous time profile of 
"classical" (very bright) TGFs is not just a function of their being observed from space but is intrinsic even when they are observed at close quarters.

The RREA feedback process provides a ceiling on the steady electric field that can be maintained in thunderstorms \citep{dwyer03limit}; beyond
this field, feedback will grow until it produces a current that cancels the thunderstorm's charging rate, producing a steady glow of bremsstrahlung gamma-rays in the process.  There is observational evidence that this field limit is, indeed, close to the maximum field observed in thunderclouds \citep{marshall95}, and glows
have been observed from aircraft \citep{mccarthy85,kelley15}, balloons \citep{eack96balloon,eack00}, and the ground \citep{brunetti,torii02}, where they are sometimes referred to as long bursts \citep{tsuchiya07}, or thunderstorm ground enhancements (TGEs) \citep{chili13}.
The faintest glows may not involve RREA, being instead due to acceleration of cosmic-ray secondary electrons and positrons without avalanching in a process called modification of spectrum (MOS) that produces a hard power law spectrum \citep{chili13}.  But the brightest glows, sometimes observed
from within the electron avalanche itself, show the characteristic spectrum of RREA and can be strong enough to plausibly represent a discharge mechanism comparable in importance to lightning \citep{kelley15}.

\section{Observations}

Our observations are part of a long-running campaign to study winter lightning
on the western coast of Honshu, Japan in the town of Uchinada, Ishikawa Prefecture.  The instrumentation there includes a 30,000 frame-per-second camera (NAC MEMRECAM GX-8) permanently 
viewing a lone wind turbine and a lightning protection tower adjacent to it
\citep{wang12}, electric field mills, and recently the Gamma-ray Observations During Overhead Thunderstorms (GODOT) instrument, a suite of gamma-ray detectors that have also proved sensitive to neutrons arriving in the aftermath of a TGF; this result and a discussion of the instrumentation are given in \citet{bowers17}.

That paper examined a TGF that occurred on 3 December 2015 in connection with a self-initiated upward leader from the lightning protection tower at 20:20:29~UT.  We showed that one of the GODOT detectors, a cylindrical plastic scintillator of 12.7~cm diameter and 12.7~cm length, detected a signal consistent with gamma-rays of energy 2.2~MeV produced in the detector itself by the capture of thermal neutrons by hydrogen nuclei.  Monte Carlo simulations showed that the number of neutrons and the characteristic timescale of their decay were consistent with a TGF about as bright as those seen from space, with the same gamma-ray spectrum, aimed downwards at the ground from an altitude of 1.0~km above the tower.  This altitude is consistent with the typical altitude of the main negative charge center in winter thunderstorms in this region \citep{saito09}.  A similar TGF event at another site on the Japan Sea coast was later seen by \citet{enoto17}, and a similar conclusion reached, with the additional and conclusive observation of the $\beta+$ decay of the radioactive nuclei left behind by the neutron photoproduction.

In this paper, we revisit the TGF of 3 December 2015 and its associated upward lightning in order to compare it to another upward flash that occurred almost exactly four hours earlier, at 16:20:28~UT on the same day (01:20:28 local time).  GODOT, the high-speed camera, and the field mills all had coverage
of both events.  They are the only two upward leaders observed by both instruments that were spontaneous, and thus make up a complete if limited data
sample (a third upward leader occurred on 13 January 2016, but it was in response to intracloud processes occurring overhead, and had a very different
character, being highly branched and much slower than the spontaneous leaders
discussed here).
Both events initiated at the
lightning protection tower, and both appeared to proceed directly to the cloud, producing an initial continuous current (ICC) pulse without branching within the field of view of the camera
(ICC is the term for the surge of current following the connection of an upward leader with a cloud,
analogous to an M-component or return stroke, see e.g. \citet{wang99}).
However, while the flash at 20:20:29~UT created a TGF so bright that GODOT's detectors were briefly paralyzed by the gamma-rays, coming back to life to show a flood of events related to the neutron afterglow, the flash at 16:20:28~UT produced no detectable perturbation in the ambient background level of radiation detected by GODOT.  In the following sections, we describe the video, electric field, and gamma-ray observations in more detail, and discuss a key difference in the gamma-ray data that we suspect can explain why two upward leaders so similar had such different results.

\subsection{Field mill and camera observations}\label{camera}

\begin{figure}[ht]
\centering
\includegraphics[width=35pc]{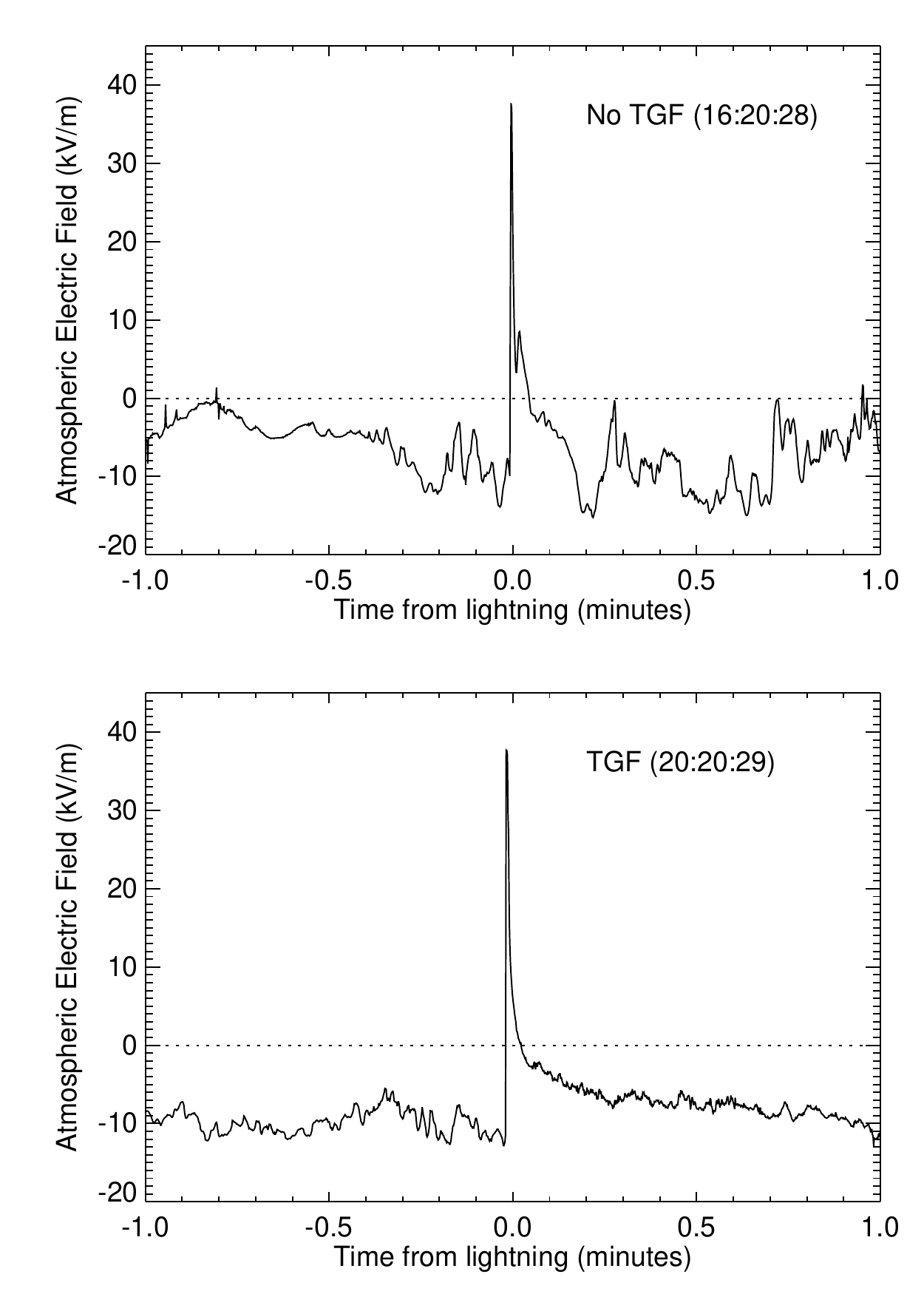}
\caption{
Electric field mill data (atmospheric electricity convention) for the
flashes with no TGF (top) and with a TGF (bottom).
}
\label{fieldmill}
\end{figure}

The field mill data (Figure~\ref{fieldmill}) from both flashes indicate that at the time the leader was 
launched from the tower, the
electric field at the ground was upward in the physics convention and
negative in the atmospheric electricity convention (shown in the
figure), meaning that a positive
leader would more easily propagate upward than a negative one. 
The charge structure in winter thunderstorms in this part of Japan can be very complicated, although
always compact and low in altitude \citep{wang18}, but the field sign
during both events is consistent
with a strong main negative charge center overhead and no
strong lower positive region, or at least not one that is very close horizontally. 
The total charge transfer during each flash, as indicated by
the field change at each of the flashes in
Figure~\ref{fieldmill}, was in the sense of bringing negative charge downwards, consistent with a -CG flash initiated by an
upward positive leader.  

\begin{figure}[ht]
\centering
\includegraphics[width=35pc]{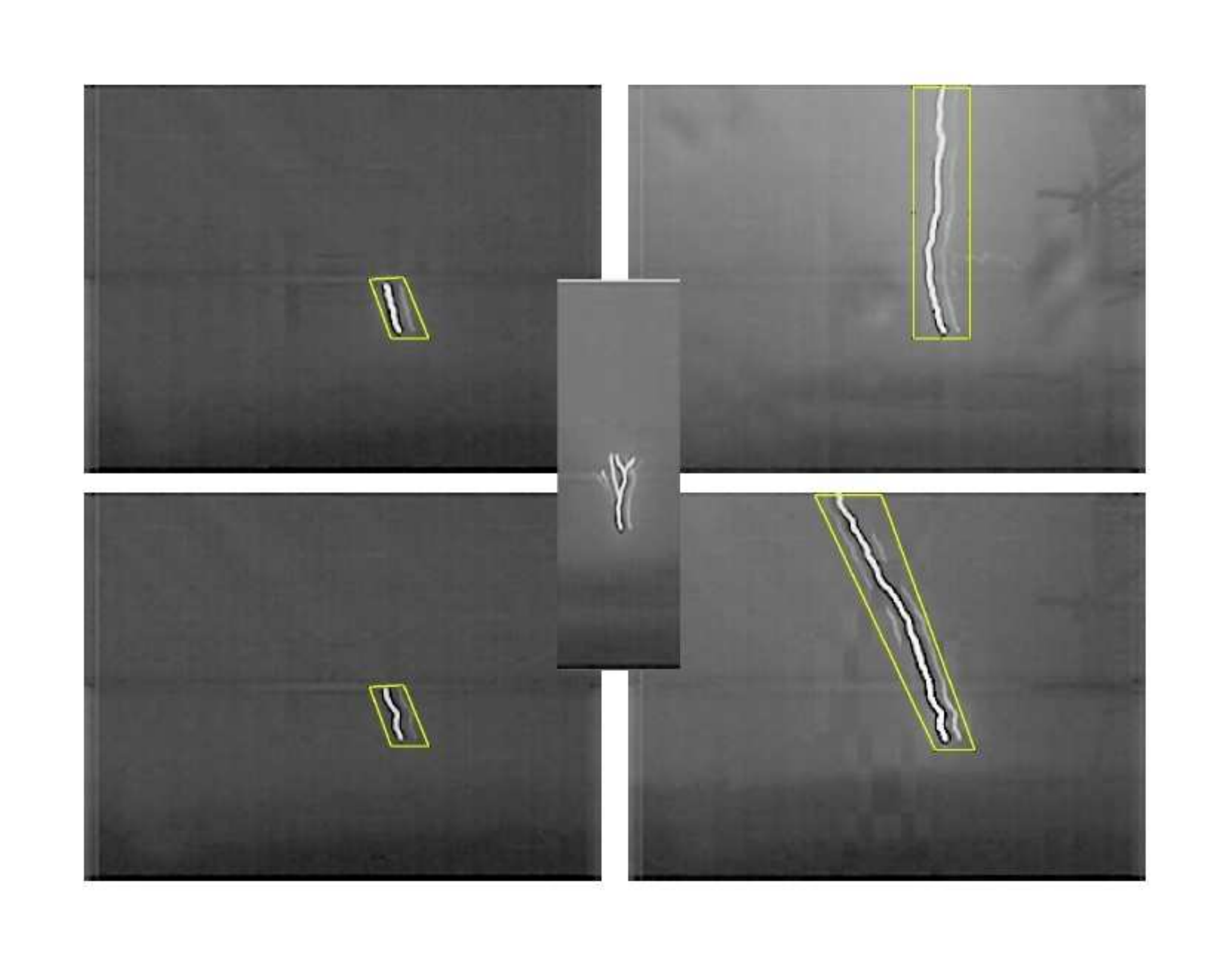}
\caption{Stills from the high-speed camera video for the gamma-ray quiet flash of 16:20:28~UT (top)
and the TGF-producing flash of 20:20:29~UT (bottom).  Left: leader development at the moment the
leader reaches the middle of the camera frame for each flash.  Right: the channel
after partially cooling from the ICC.  Inset: Still from the beginning stages of a negative upward leader at the same site from 13 January 2016 at 15:17:07.389UT, showing the
branching more typical of negative leaders. }
\label{lightning}
\end{figure}

\begin{figure}[ht]
\centering
\includegraphics[width=25pc]{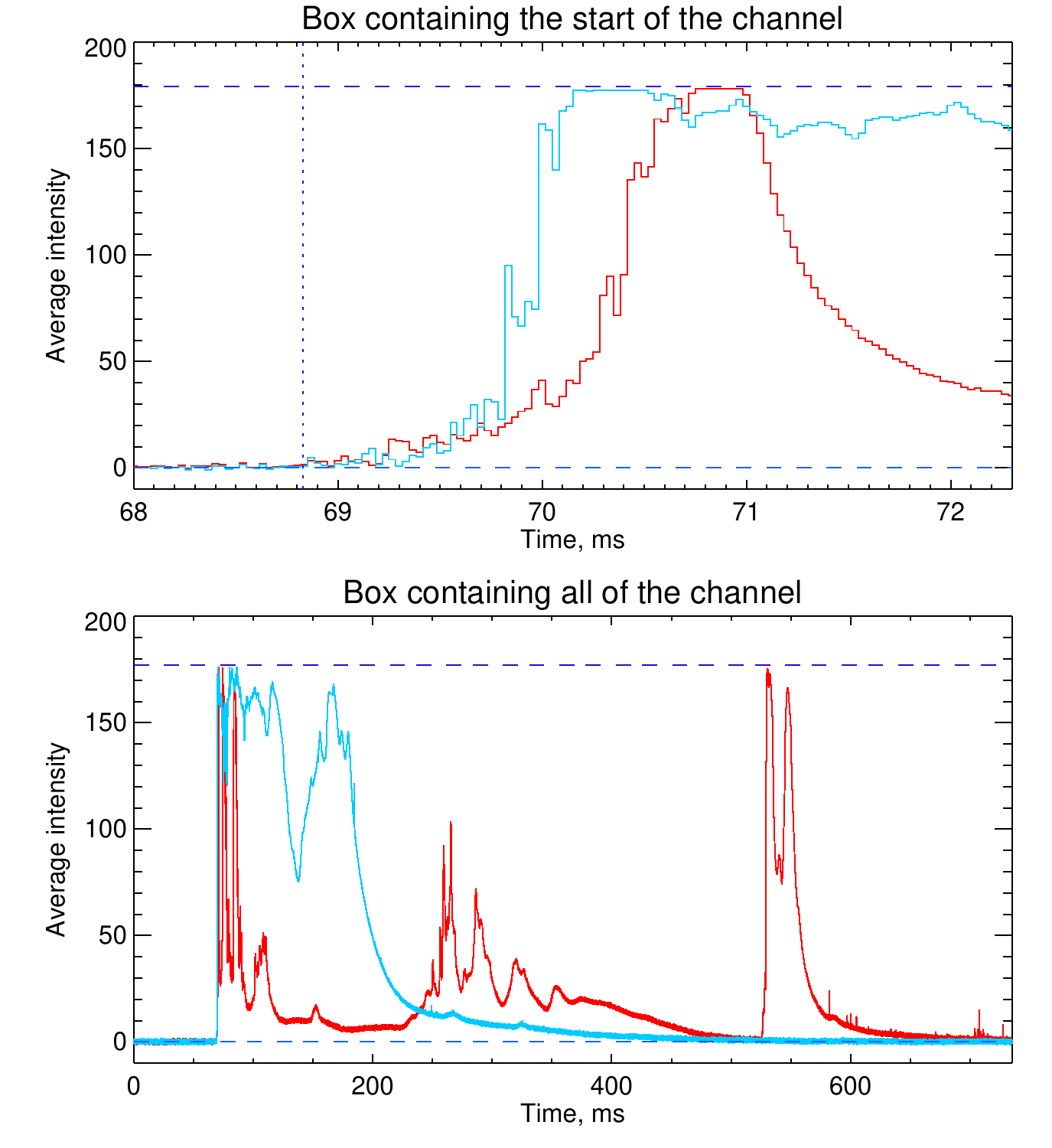}
\caption{A) Initial leader ascent and ICC shown in average pixel intensity (ADC units)
within the small boxes defined at the left of Figure~\ref{lightning}.  Background 
has been subtracted. B) Average pixel intensity in the larger boxes defined at the right
of Figure~\ref{lightning} over the evolution of the flashes.  Red curves: the TGF-producing
flash of 20:20:29~UT.  Blue curves: the gamma-ray quiet flash of 16:20:28~UT.  The horizontal dashed lines show the background and saturation levels; the vertical dotted line in the top
panel shows the point at which both leaders are first visible, defining the alignment
of the traces.}
\label{rising}
\end{figure}

Figure~\ref{lightning} shows video stills from the evolution of both flashes: one at similar stages in the ascent of the upward leader, and one taken as the hot channel from the ICC had begun to cool.  In both cases, the channel shows no
branching and ascends quickly.  The inset in Figure~\ref{lightning} shows a negative upward leader from the same
site that was part of a flash studied in great detail by the LIVE
lightning mapping array on 13 January 15:17:07.389UT (unfortunately,
LIVE data are not available for the  flashes we report on here).
This negative leader, which began in the middle of a long series
of in-cloud processes, is significantly branched.  Negative leaders
generally show more branching than positive ones, although positive
leaders can branch as well; but overall, this comparison between leaders
from the same site is at least consistent with the other arguments
that both the events of 3 December 2015 are positive leaders. 

There are many other upward leaders initiated later in the 
flashes of 3 December 2015, 
from other spots that probably include the turbine blades and the 
towers of a nearby bridge.  These leaders are
branched, slow, and tortuous, and most never connect
to the cloud or show subsequent brightening of the channel. 

Figure~\ref{rising} shows the evolution in overall brightness of both flashes, both during
the initial ascent of the leader and ICC (top panel) and during the entire
flash as captured by the camera (bottom panel).  For the top panel, only the pixels in the 
small yellow box (left panels of Figure~\ref{lightning}) are used, while in the lower panel the whole visible portion of the channel is used, which requires the definition of different
boxes for each flash (right panels of Figure~\ref{lightning}).  In this video ({\tt .avi} format), pixels saturate at a raw value of 254~analog-to-digital converter (ADC) units, and the background brightness before the flashes, which is subtracted from the average brightnesses in Figure~\ref{rising}, 
has an intensity level of around 75~ADC units.  During the ICC, when the channels are hottest, all the pixels within the boxes are completely saturated (horizontal purple dashed lines in Figure~\ref{rising}).  The absolute timing of the TGF data in GODOT is uncertain enough that we cannot
be sure whether it begins just before or just after the saturation (ICC) in the red
trace in Figure~\ref{rising}A, but we know that it is associated with this stage of the flash and
not any of the later brightenings shown in Figure~\ref{rising}B.  

The leader without the TGF (shown in blue in Figure~\ref{rising}) rises more quickly in both height and intensity, which could indicate either a more strongly charged leader tip or a stronger ambient
field. The time from the first appearance of
a signal above the background to complete saturation of the small box (top of Figure~\ref{rising}) --
probably coinciding with the ICC -- is 1.3~ms (39 frames) for the gamma-quiet flash and 
1.9~ms (57 frames) for the TGF flash.  The leader in the gamma-quiet event also rises more quickly in
altitude, taking $\sim 0.5$~ms to climb from the center line of the camera frame to the top of the
frame as opposed to $\sim 1.3$~ms for the leader that produced the TGF.  At the distance
from the camera to the tower, this vertical distance corresponds to 173~m, giving minimum speeds of 
$1.3\times10^5$m/s for the TGF leader (which appears essentially vertical) and $3.9\times10^5$m/s for the gamma-quiet leader (tilted at approximately 23$^{\circ}$).  These speeds are close to the upper and
lower ends of the range of speeds found by \citet{wang12} for five
earlier self-triggered, upward positive leaders at this site.  At another, similar
site at 190~m greater altitude,  \citet{wang12} found much higher speeds
for otherwise similar leaders, averaging around $1\times10^6$m/s.
Any additional tilt in the direction
radial from the camera would imply a higher speed, 
which is of course proportional to the inverse cosine of
the tilt angle.  This is the dominant source of uncertainty in these speed
measurements.

Finally, the gamma-quiet flash remains saturated slightly longer, and shows a much more persistent high luminosity, than is the case for the flash that produced the TGF (see the bottom panel of Figure~\ref{rising}). While these measurements do
not comprehensively characterize either flash, if anything they suggest that the
leader and ICC which did not involve a TGF were more powerful than the
ones that did, rather than the reverse.

\subsection{Gamma-ray observations}

\subsubsection{TGF and TGF limit}

\begin{figure}[ht]
\centering
  \includegraphics[width=25pc]{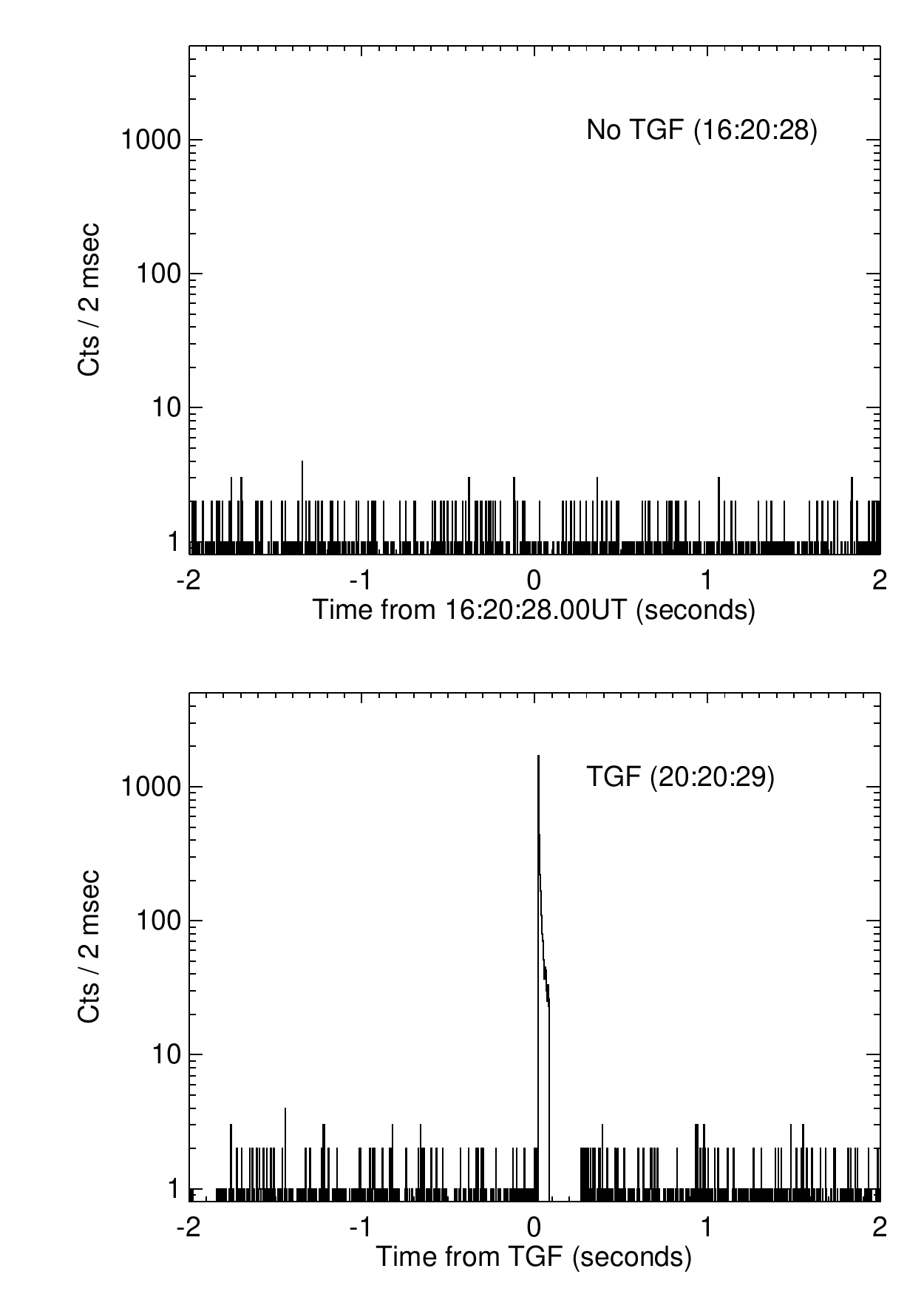}
\caption{Counts in the large plastic scintillator per 2~ms time interval for the gamma-ray quiet
flash (top) and the TGF-producing flash (bottom).  The TGF gamma-rays themselves paralyzed the
instrument but would have reached a value of $\sim 5.0 \times 10^7$ in a single time bin on this
plot.  The data gap visible in the plot is not the very short paralysis during the TGF, but rather
missing data caused by the buffer readout time required to capture the neutron-induced counts
(see \citet{bowers17}).
}
\label{tgf}
\end{figure}

The background count rate in the large plastic scintillator around the time of the gamma-ray
quiet flash of 16:20:28~UT was 179 counts per second.  Since nearly all known TGFs have a duration of less than 2~ms or less in their gamma-ray emission \citep[e.g.][]{gref09rhessi}, we searched
the data for peaks associated with this flash using that time scale.  The top panel of Figure~\ref{tgf} shows the number of counts in 2~ms intervals around the quiet flash.  At an
average rate of 0.36 counts per 2~ms bin, Poisson statistics predict 3 counts or higher in 0.6\%
of samples on average.  This level is reached in 8/2000 or 0.4\% of the samples in this figure, 
and 4 counts appears once, both consistent with the expected Poisson statistics; we therefore consider a level of 5 counts, the
lowest level that does not appear at all in this data set, to
be the upper limit for any TGF signal associated with this flash, regardless of the stage
of the flash at which it might have occurred.  With the effective
area of the large plastic scintillator to a TGF-like spectrum being 100~cm$^2$, this implies a
limit on the TGF fluence of 0.05~photons~cm$^{-2}$.

Conversely, the number of gamma-ray counts in the TGF of 20:20:29~UT can only be estimated by modeling, since the entire instrument was paralyzed by the gamma-ray emission. The extremely 
bright peak in the bottom panel of Figure~\ref{tgf} is only the response of the detector to the 
neutron afterglow, whose instantaneous flux (particles per square centimeter per second) and total fluence (particles per square centimeter) are both much fainter than those of the unrecorded TGF itself.  By simulating the production of neutrons in the atmosphere, their propagation through
the atmosphere and in the ground, and their interactions in the GODOT detector system, \citet{bowers17}
were able to get good agreement with the brightness and time evolution of the observed neutron signal in this event with a downward TGF producing $10^{17}$ gamma rays at an altitude of 1~km above the tower (see Figure~5 of \citet{bowers17}).  The gamma-ray fluence would have been $5.0\times 10^{5}$ photons~cm$^{-2}$ giving 
$5.0\times 10^{7}$ expected counts in the large plastic scintillator, all within a single
2~ms time bin.  This exceeds the counting ability of the detector system by more than four orders
of magnitude, consistent with the observed brief data gap just after the start of the TGF,
and before the neutron signal, being due to paralysis during the TGF itself (see Figure~2C of \citet{bowers17}).

Combining these two results, we find that the ratio between the modeled TGF counts in the
20:20:29~UT flash and the limit for the 16:20:28~UT flash is $5.0\times 10^{7} / 5$ or a factor of 
$1.0 \times 10^7$.  This
is an extraordinary difference for two flashes that look similar in many ways, except for those
observations that point to the leader {\it without} the TGF as being perhaps somewhat more powerful.

\subsubsection{Dependence on geometry}

This limit relies on the assumption that the geometry is the same for both the point where the TGF
is created and the point where we think it could have been created in the other flash; the geometrical
factors include the horizontal displacement of the TGF origin from GODOT (300~m if the leader is perfectly vertical
from the tower), the direction of the electric field in the avalanche region, and the altitude of the
avalanche region.  Because the high-speed camera is not stereoscopic, it is possible that either flash
or both had significant tilt along the direction between the camera and the tower; highly-tilted
upward leaders have been observed at this site \citep{wang12}. The line between the tower and camera
is about 43$^{\rm \circ}$ from the line between the tower and GODOT, and is a bit over three times
the distance.  A leader that left the tower tip and headed out toward the sea would be moving in a
direction approximately away from both the camera and GODOT.  If the leader that
produced no TGF was highly tilted directly away from the camera (and also GODOT), or if it met the
main negative charge region at a much higher altitude, then we would be
less sensitive to a TGF it might have produced, and the ratio of $1.0 \times 10^7$ between the TGF
gamma-ray flux and the upper limit from the other flash would not be as extreme.

\begin{figure}[ht]
\centering
\includegraphics[width=25pc]{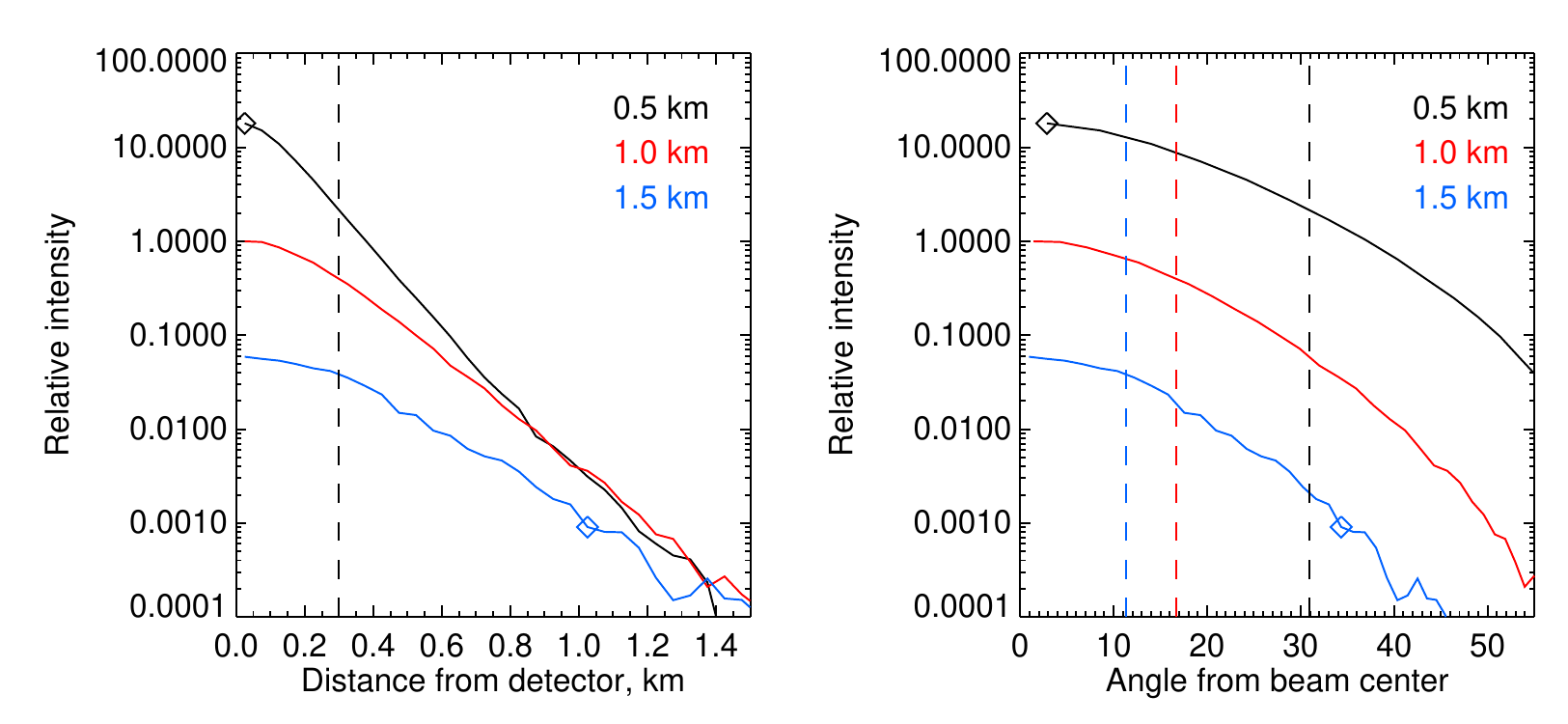}
\caption{Left: relative simulated intensity (photons per square centimeter) as a function of 
horizontal distance from the footpoint of a downward TGF at three altitudes.  The dashed line shows
the distance of GODOT from the tower.  Right: the same curves plotted with the angle from beam center
as the abscissa.  GODOT's position in this coordinate system is dependent on the TGF altitude, as shown.
The diamonds represent the extreme cases from the text that would be very favorable for the detection of
the TGF (black) and very unfavorable for the detection of a second TGF from the flash that was observed to be gamma quiet (blue).
}
\label{geant}
\end{figure}

There are two arguments that tend to reduce, though not eliminate, this concern: first, that at least
in the plane of the camera's field, the gamma-quiet leader that appears almost perfectly vertical (Figure~\ref{lightning}) so that if it is, in fact, highly tilted, that tilt was serendipitously oriented
rather precisely along the axis from the camera to the tower.  Second, we assume that these upward leaders
have a preference for following the direction of the electric field \citep{wang12}; thus, the field at the cloud base and
the direction of the leader propagation might prefer to be somewhat aligned.  In that case, the direction
of the RREAs, and therefore the TGF gammas, would be preferentially in the direction back toward the 
tower instead of straight downwards.   This would put the tower (and GODOT, which at 300~m distance
can be considered close to it), close to the beaming direction of the TGF even if it was produced
further away.

Nonetheless, in order to examine something close to a worst-case scenario, we performed Monte Carlo
simulations of the propagation of TGF gamma-rays from the source point to the ground using the GEANT4
package \citep{geant}.  The angular, altitude, and energy distributions of the gamma-rays for input to GEANT4 were calculated using the Relativistic Electron Avalanche Model (REAM) code \citep{dwyer03limit}.
The angular distributions thus obtained are shown in \citet{hazelton09}, but are broadened, particularly
at lower energies, by Compton scattering in the atmosphere on the way to the ground.  
This angular distribution was centered on the downward direction when the photons were initialized.  Gamma-rays were
collected as they reached the ground in the simulation in annuli of 50~m width around the point directly below the TGF.  Figure~\ref{geant} (left) shows the relative intensity at the ground for various radial distances 
from the center of a downward TGF produced over a range of altitudes appropriate for the bottom of the
negative charge center for upward flashes in winter thunderstorms in this region \citep{saito09}.  The 
right hand panel shows the same intensity patterns transformed into an angular coordinate system, 
which better isolates the change in luminosity of the pattern due to the
different source altitudes, since the effect of the simple expansion of the conical beam is removed.
From these simulations we conclude that even in very unfavorable scenarios, differences in altitude
and horizontal offset cannot account for the observed ratio of at least $1.0 \times 10^{7}$ in flux.  To take an
extreme case, let us say that the TGF of 20:20:29~UT was directly overhead of GODOT (not the tower where it originated) and as low as 0.5~km, maximizing GODOT's sensitivity to it.  Then, for the gamma-quiet leader, let us take the corresponding altitude (the altitude where the leader approaches the negative charge center) as 1.5~km, three times higher, making us much less sensitive to gamma-rays produced there, and let us further disadvantage this event by placing it at a horizontal offset of 1.0~km.  These cases are shown as black and blue diamonds, respectively, in Figure~\ref{geant}, and they are a factor
of 20,000 different in fluence on the ground at GODOT; so even this extremely unfavorable geometry would retain a factor of 500 ratio between the intrinsic brightness of the observed TGF and the limit
from the flash at 16:20:28~UT.

While we have not run simulations with a tilted beam, the effect of a tilt can
be easily estimated by simple geometry and reference to Figure~\ref{geant}.  The intensity of the TGF radiation field is primarily a
function of the angle relative to the beam center and the total distance through
the air from the source to the detector.  Thus, the right hand panel of Figure~4
can be used to estimate the intensity in the case of titled beams by taking the
x-axis as referring to the angle of the source/detector line relative to the beam
center, and using the source/detector distance instead of the source altitude to
interpolate between the curves.

\subsubsection{Glows}

\begin{figure}[ht]
\centering
\includegraphics[width=25pc]{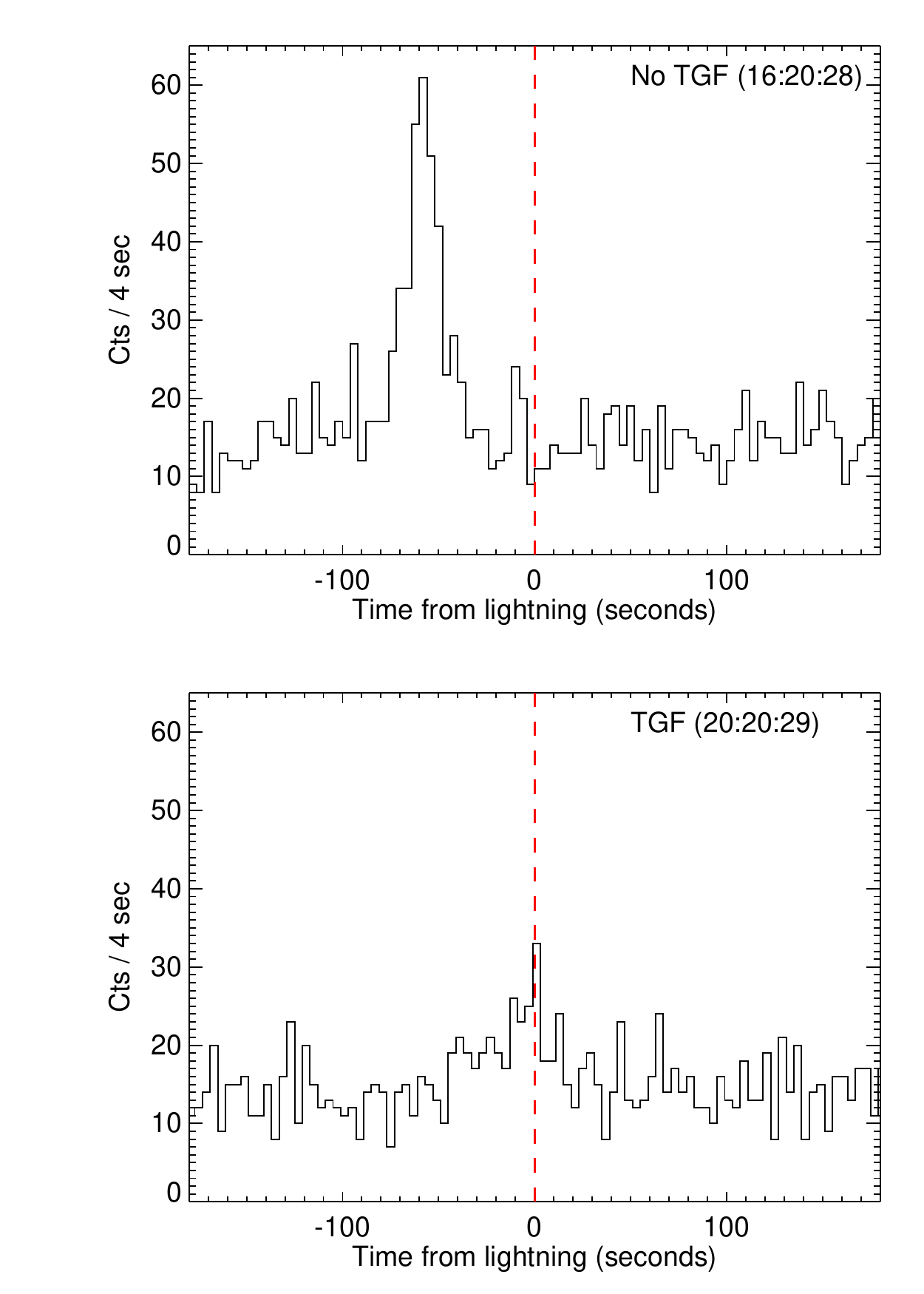}
\caption{Counts per 4-second interval above 2.8 MeV in the large NaI detector.  The gamma-ray
quiet flash (dashed red line, top panel) occurred well after a glow had ended, but the TGF-producing
flash (dashed red line, bottom panel) occurred at the peak of a modest glow, and may have terminated it.}
\label{glow}
\end{figure}

Observations of the gamma-ray rates before and after these two flashes provide us a possible solution
to this paradox.  Figure~\ref{glow} shows the gamma-ray count rate above 2.8~MeV in the large
NaI detector in 4~s time bins around each event.  This energy threshold allows for sensitive
detection of glows because it is above most natural emissions from radioactivity, and the
ambient background count rate is dominated by cosmic rays, which are a much weaker background at
sea level.  The NaI detector is the best component of GODOT for studying glows, which do not
produce count rates that challenge its limited throughput.  NaI not only has more mass to
stop high-energy gammas than the plastic scintillators, its higher atomic numbers produce more
photoelectric absorption, so more of the interacting photons leave 100\% of
their energy in the detector, allowing better spectroscopy.

The count rates in Figure~\ref{glow} show that while there was a glow in progress shortly before the flash that
produced no TGF, it was over long before the flash occurred.  On the other hand, the TGF-producing
flash occurred in the middle of a glow event in the cloud above, during the brightest
4-second interval of the glow. Thus, we find that the large-scale
thundercloud field was higher during the time of the TGF.  Since there was no 
evidence to indicate that the leader in the case of the TGF was faster or more
highly charged, and indeed the hints are to the contrary (section \ref{camera}),
this pair of events is at least suggestive of the idea that the field aloft may be
more significant in determining whether a TGF will take place than the details of the leader.

The counts from the TGF
itself have been removed from the bottom panel of Figure~\ref{glow}, although there were very
few in the NaI detector outside its period of paralysis.  In addition, the time spent in paralysis has
been excluded for purposes of calculating the glow count rate.  
The glow is statistically significant; using a time interval $\pm 40$s around the TGF, we find 367 counts where the background derived from 800~s of
nearby data would give 274.8 counts, giving an excess of 92.2
counts or 5.6$\sigma$.  But the excess after
the time bin containing TGF is not significant at all, so we cannot conclude whether the glow continued after the TGF or
was terminated by the movement of charge associated with the lightning and/or the TGF.

Geometrical factors 
(the height of the high-field region and its lateral distance from GODOT) being taken
as equal, the implication of these observations is that the in-cloud electric field was higher
for the flash that produced the TGF, and therefore that the production of a TGF may depend more
sensitively on the maximum in-cloud field that a leader encounters than on characteristics
of the leader itself. 

\section{Discussion}

\subsection{TGF frequency}

We consider the implications of these observations for the questions of how common TGFs are,
and whether they have a continuous distribution of brightnesses embracing bright and weak
events, or a minimum threshold luminosity.

First, since the two events reported here are the only two spontaneously initiated upward leaders
seen by both the high-speed camera and GODOT, we find that 1 out of 2 observed leaders produced
a TGF.  If we assume a flat prior (no prior knowledge) of the true fraction, $f$, of upward leaders
that will produce a TGF, then applying Bayes's theorem to the binomial probability $P(\frac{1}{2} | f)$
yields a probability distribution for $f$ of $P(f | \frac{1}{2}) = 6f(f-1)$ \citep[][pg. 76]{gregory}, giving a 95\% confidence interval for $f$ of 0.094 to 0.906.  

While this is of course a wide range, it is worth comparing it to 
what we know about the fraction of lightning producing upward TGFs as seen from space. 
The fraction producing {\it observed} TGFs is less than 1\% and probably closer to 0.1\%  \citep{fuschino11,ostg12,briggs13,tierney13}. Our recent analysis of the sum of gamma-ray emission observed by the RHESSI spacecraft while
flying over many lightning flashes \citep{smith16} demonstrated that there is remarkably little
gamma-ray emission coming from lightning as a whole that is not already accounted for by TGFs bright
enough to trigger the spacecraft's detection algorithms.
This allowed us to conclude that for most simple brightness distributions, the {\it total}
fraction of TGFs, whether individually observable or too faint, was still $<1$\%.  The binomial probability
of seeing 1 or more TGFs in two trials with $f=0.01$ is 0.020, so we can say with 98\% confidence that these two occurrence rates are inconsistent.  We do not interpret this
as a matter of an error, but rather that upward leaders from this tower in winter
thunderstorms are more effective at producing TGFs than lightning in general.  Of course
this conclusion will be much stronger when more similar observations are made at this and
similar sites, but it is essential to keep track of all events that do not produce TGFs in
addition to noticing and publishing the ones that do.

\subsection{TGF luminosity distribution}

\citet{bowers17} showed via simulations that the TGF observed at 20:28:29~UT would have 
involved on the order of $10^{17}$ initial gamma rays (above 1~MeV) if it originated at an altitude of 
1~km, typical of the main negative charge center in winter thunderstorms in this region.  This
is comparable to what is expected of some of the fainter TGFs observed from orbit, although in
both cases the result is sensitive to the choice of altitude.  It is subtle but significant that
with only a single TGF observation at this site, we find a luminosity comparable to those seen
at great distance from space.  As pointed out above, at GODOT's position relative to the TGF,
we would have been sensitive enough to see a TGF even 10,000,000 times fainter than the one we
saw.  Thus if the luminosity distribution of TGFs was a falling power law down to luminosities
much lower than  $10^{17}$ gamma rays, the probability that the first one observed would be 
fully as bright as those seen from space should be very small.  

To pick a simple example, we take the distribution of observed TGF luminosities from space,
a power law with the differential number spectrum with fluence $\frac{dN}{dF}$ going as approximately 
$F^{-2.3}$ \citep{ostg12,tierney13,marisaldi14}, and let us assume that it has an abrupt cutoff
below which there are no TGFs.  Of course this model may not be applicable to our tower observations
for two reasons: first, it's a different population of TGFs with a different kind of causative
lightning; and, second, the observed TGF fluence distribution is complicated by the distribution of 
distance from the spacecraft to the TGF \citep[][Figure~3]{smith16}, and thus is not really an
intrinsic luminosity distribution for TGFs.  Still, this model, which was one of the cases
examined in \citet{smith16}, gives an example of the significance of not having yet seen a weak
TGF, even out of a sample of only two observations.  If we assume this distribution must integrate
to a total probability of $P_0 =0.5$, which is our observed TGF fraction (1 out of 2 observations), 
the power law index is $\alpha = 2.3$, and the number of counts in GODOT corresponding to the
cutoff fluence is $C_0$, then the normalized differential probability distribution of the number
of observed counts $C$ is
\begin{equation}
\frac{dP}{dC} = \left\{
\begin{array}{ll}
\frac{P_0}{C_0} (\alpha - 1) \left( \frac{C}{C_0}\right) ^{-\alpha}  & C \geq C_0\\
0 & C < C_0\\
\end{array}
\right.
\end{equation}
Given that we infer a number of counts $X = 5.0\times 10^7$ for the TGF we saw at Uchinada, 
the probability of observing
something at least this bright, as a function of $C_0$, is the integral of this function
above $X$, or $P(\geq X) = P_0 \left(\frac{X}{C_0}\right)^{1-\alpha}$.  The probability of seeing
something this bright {\it for the first real TGF observed} simply removes the factor of $P_0$.  For 
a 95\% confidence limit, we set $P(\geq X) = 0.95$ and find $C_0 = 8.5\times10^6$, i.e. under this
model, with 95\% confidence, there are no TGFs a factor of 6 or more fainter than the one observed.
As it happens, the 95\% confidence lower limit derived for the cutoff in
same model in the stacking analysis of RHESSI TGFs (see Figure 11a in \citet{smith16}) was about 2 counts in RHESSI, or
about $\frac{1}{6}$ the number of counts in a weak but individually detectable RHESSI TGF,
also corresponding to roughly $10^{17}$ gammas at the source \citep{dwyersmith05}. This
close agreement is almost certainly coincidental, since we have no reason to expect
that the luminosity distribution function should be the same for these very different TGF causative 
mechanisms and environments.  But the qualitative conclusion, that there is not
a large population of faint TGFs within a couple of orders of magnitude of the ones we know of, is
strong in both cases.

\subsection{Theoretical implications}

One appealing model that has been proposed for upward TGFs is that they are simply the result of a leader producing sub-relativistic electrons through cold runaway encountering a large region of field in the cloud that is high enough to support relativistic runaway \citep{moss06,dwyer08,dwyer12tgftheory}.  Over a wide range of high thunderstorm charging rates, the thundercloud field can be kept near a single value, the threshold for RREA feedback \citep{dwyer03limit}, by the feedback mechanism itself.  At this field value, RREA provides an avalanche multiplication factor of $\sim 10^4$ \citep{dwyer08} to external seeds, such as those from a leader.  
At lower fields, below the RREA threshold for seed electrons, there is no electron multiplication at all, although
gamma-ray yields per electron can be enhanced by higher electron energies.
The exact ratio between the field value just before RREA begins (no multiplication) and the field value at which feedback is limiting further increase ($10^4$ multiplication) is dependent on the energy of the seed particles, the length of the avalanche region (i.e. potential as distinct from field), and the
field geometry (purely parallel vs. diverging field lines).  But if the difference between these
field thresholds is small for typical thunderstorm scenarios, there could be a natural explanation
for a TGF luminosity threshold: in-cloud field values at the point where a leader approaches a charge center might vary greatly, and would then be statistically likely to fall either in the no-RREA or maximum-RREA ranges, and unlikely to fall within the small range between.  As a possible mechanism
for a threshold effect in {\it upward} TGFs, this is a useful target for further modeling.

However, in the current case, the leader for the TGF discussed here and in \citet{bowers17} was a positive leader approaching a negative region of charge, the opposite of the case for upward TGFs.
Since negative electrons are the easily accelerated charge carriers in either case, the situations
are not symmetrical.  A negative leader approaching a positive charge region can easily be
imagined to inject mildly relativistic seed electrons forward into the avalanche region,
producing a TGF well before it reaches the charge region
it is approaching.  The upward positive leader, however, as it 
approaches the high-field region, accelerates its seed electrons away from that region down into its channel. 
\citet{moss06} considered positive leaders unlikely sources of seed electrons both because they propagate
counter to the leader tip and because the positive streamer threshold field is lower than that for
negative streamers.
Thus if these downward TGFs proceed from leader-generated seeds at all, the mechanism is likely
to be more complex.  For example, the TGF may require a downward negative leader to form in response to the approaching positive leader from the ground, in analogy to the attachment process when a leader
approaches the ground.  If the seeds come directly from the upward leader, on the
other hand, the TGF may have to 
occur after the leader has penetrated the negative charge region, just at the start of the ICC, because the seeds must have most of the available potential below them to produce avalanches.
In this case, the seeds may come from the bottom of the space stem formed during the stepping process just above the leader, or may be produced by bremsstrahlung x-rays that can escape the leader tip
when electrons are accelerated into it.

If the TGF seeds are normally dominated by the feedback process and not mostly contributed by the leader anyway \citep{dwyer12tgftheory}, the asymmetry vanishes, and the leader's function in triggering the
TGF is simply to abruptly push the local field well above the feedback threshold, resulting in
explosive feedback until the TGF-related discharge brings the field below the threshold again.  While
this mechanism allows positive and negative leaders to be equally effective in triggering TGFs,
it may not present as simple a picture in terms of what initial field is required for a TGF; that
threshold might no longer be independent of the charge on the leader as suggested at the start
of this section. 

One might imagine that even a cloud not currently experiencing RREA and a glow might still produce
a TGF if the leader charge was high enough to push the local field at once past both the minimum
RREA and feedback-limited field thresholds.  Even the glow we observed could conceivably have been due
to MOS rather than RREA in the feedback-dominated scenario, since it wouldn't have to provide full RREA
avalanching until the field was enhanced by the leader.  While our instrumentation is not sensitive enough to measure the spectral difference between MOS and RREA glows, which have similar spectral slopes below about 10~MeV and only start to look significantly 
different when the RREA exponential cutoff becomes obvious above that energy
\citep{chili13}, the expected ambient atmospheric density of relativistic particles sets a limit on the possible intensity of a MOS glow.  With a good model of that density, and the altitude and lateral extent of the glow region, it might be possible in future work to declare that any given glow must have been due to
RREA simply because it is sufficiently bright.  For the moment, we confine ourselves to the observation that
whether the RREA threshold was already crossed in the TGF-producing storm when the upward leader came, at any rate the field was probably higher in that storm at that moment than it was in the earlier storm at the moment of the upward leader that was gamma-quiet.

We look forward to future observations at this site and similar sites in Japan in which a greater number of upward leaders in winter thunderstorms can be studied to see if their TGF luminosity (or upper limit) follows the
preliminary "all-or-nothing" pattern implied by these two events and our statistical study from space
\citep{smith16}.  We also look forward to modeling studies that will combine field and feedback calculations
as a function of charging rate with estimates of leader production of seed electrons to test the
expectation of a TGF luminosity threshold in the leader-seed-dominated model.

\acknowledgments
The authors thank Takuma Tsuruda and Shusa Takahashi for their work
with the field mill instruments.
JRD and DMS acknowledge the support of National Science Foundation awards AGS-1612466, AGS-1160226, and AGS-1613028.  The data and code underlying
the plots presented here are available at 
{\tt http://research-archive.scipp.ucsc.edu/godot/}.

\listofchanges

\end{document}